\documentclass[aps,pra,superscriptaddress,twocolumn,10pt]{revtex4-2}
\usepackage{dcolumn}
\usepackage[utf8]{inputenc}
\usepackage{amsmath}
\usepackage{amsfonts}
\usepackage{amssymb}
\usepackage{graphicx}
\usepackage{hyperref}
\usepackage{xcolor}
\usepackage{dsfont}

\newcommand{\oper}[1]{\mathbf{\mathsf{#1}}}

\begin{document}

\title{Second-order moment equivalence of twisted Gaussian Schell model beams and orbital angular momentum eigenmodes}
\author{T. Ferreira}
\affiliation{Instituto de F\'{\i}sica, Universidade Federal Fluminense, Brazil}
\affiliation{Departamento de F\'{\i}sica, Universidad de Concepci\'on, 160-C Concepci\'on, Chile}
\affiliation{Departamento de F\'isica, Universidad del B\'io-B\'io, Collao 1202, 5-C Concepci\'on, Chile}
\author{G. Santos}
\affiliation{Departamento de F\'{\i}sica, Universidad de Concepci\'on, 160-C Concepci\'on, Chile}
\affiliation{ANID – Millennium Science Initiative Program – Millennium Institute for Research in Optics, Universidad de Concepci\'on, 160-C Concepci\'on, Chile}
\author{S. Ayala}
\affiliation{Departamento de F\'{\i}sica, Universidad de Concepci\'on, 160-C Concepci\'on, Chile}
\affiliation{ANID – Millennium Science Initiative Program – Millennium Institute for Research in Optics, Universidad de Concepci\'on, 160-C Concepci\'on, Chile}
\author{Lucas Hutter}
\affiliation{Instituto de F\'{\i}sica, Universidade Federal do Rio de Janeiro, Caixa Postal 68528, Rio de Janeiro, RJ 21941-972, Brazil}
\affiliation{Instituto de F\'{\i}sica, Universidade Federal Fluminense, Brazil}

\author{E. S. G\'omez}
\affiliation{Departamento de F\'{\i}sica, Universidad de Concepci\'on, 160-C Concepci\'on, Chile}
\author{G. Lima}
\affiliation{Departamento de F\'{\i}sica, Universidad de Concepci\'on, 160-C Concepci\'on, Chile}
\affiliation{ANID – Millennium Science Initiative Program – Millennium Institute for Research in Optics, Universidad de Concepci\'on, 160-C Concepci\'on, Chile}
\author{G.~Ca\~{n}as}
\affiliation{Departamento de F\'isica, Universidad del B\'io-B\'io, Collao 1202, 5-C Concepci\'on, Chile}
\author{S. P. Walborn}
\affiliation{Departamento de F\'{\i}sica, Universidad de Concepci\'on, 160-C Concepci\'on, Chile}
\affiliation{ANID – Millennium Science Initiative Program – Millennium Institute for Research in Optics, Universidad de Concepci\'on, 160-C Concepci\'on, Chile}

\begin{abstract}
{{We show that the covariance matrix of any cylindrically symmetric coherent
orbital angular momentum (OAM) eigenmode with quantum number $\ell$ takes a universal
form depending only on $\langle r^2\rangle$, $\langle k_r^2\rangle$, and $\ell$,
independently of the radial profile, and that this form is identical to the
covariance matrix of a twisted Gaussian Schell-model (TGSM) beam.} More specifically,  both matrices share the same pattern of zero
and nonzero entries, with the off-diagonal blocks proportional to $\ell$ and the TGSM
twist parameter $u$, respectively. This result holds for an arbitrary radial profile and provides direct term-by-term identification of
parameters between the two sets of beams. We work out the correspondence in detail for three important
families: Laguerre--Gaussian (LG), Bessel--Gaussian, and perfect vortex beams (PVBs), and
derive the conditions under which each coherent OAM mode maps onto a physically
realizable TGSM beam. {Because the covariance matrix governs second-moment evolution under arbitrary
ABCD (symplectic) transformations, any two beams sharing the same covariance matrix
are second-order indistinguishable at every propagation plane.  In particular, the
matched TGSM and coherent OAM beams share identical beam-width evolution,
far-field divergence, and $M^2$ beam-quality factor.} In particular, the
well-developed TGSM propagation toolbox applies directly to the second-order moment evolution of the three coherent
families. We further show that within each beam family the covariance matrix uniquely
determines the beam parameters, with exact uniqueness established for LG modes.
Additional results include cross-family second-moment equivalence conditions and
a proof that PVB modes form a complete orthonormal basis in the limit $w\to 0$.}
\end{abstract}

%\pacs{05.45.Yv, 03.75.Lm, 42.65.Tg}
\maketitle
\section{Introduction}
%\spw{Check references 1-13 to see if correct}
The early 1990s witnessed two major and largely independent developments in the study of structured optical fields.  
The first was the renewed recognition that paraxial optical beams may carry
quantized orbital angular momentum (OAM) given by $\ell $ per photon, following the landmark work of
Allen \emph{et al.} \cite{allen1992}, which identified the Laguerre--Gaussian {(LG)}
modes as eigenfunctions of the OAM operator $\oper{L}_z$.  
Closely related families such as Bessel--Gaussian (BG) beams and other vortex-type solutions were shown
to carry similar azimuthal structure \cite{bagini1995,mcgloin2005}.
These beams became tools of central importance in classical optics, with applications in optical manipulation, holography,
microscopy, and classical multiplexing in high-dimensional OAM spaces, and the study of topological defects and singular optics 
\cite{molina2007,willner2015,padgett2017,roadmap2017,yao2011}. In quantum optics, OAM modes have been used for quantum information encoding in
high-dimensional Hilbert spaces \cite{malik2026highdimensionalquantumphotonicsroadmap},
quantum state engineering \cite{walborn04a}, metrology \cite{dambrosio13b}.
\par

The second was the generalization of classical coherence theory to include
\emph{twisted Gaussian Schell-model} (TGSM) beams, introduced as an extension
of the well-known Gaussian Schell-model formulation of partially coherent
fields \cite{gori1983,starikov1982,simon93}.  
Classical Gaussian Schell-model beams are defined by a Gaussian intensity
profile and a Gaussian spatial coherence function. TGSM beams extend this formulation by incorporating a
nontrivial \emph{twist} produced by a phase term coupling the transverse coordinates in the mutual coherence function. TGSM beams represent the first
class of partially coherent fields whose second-order statistics naturally
encode nonzero orbital angular momentum, even in the absence
of a coherent wavefront vortex.  
They provided a pathway for generalizing concepts such as optical vorticity, phase-space rotation, and coherence-induced beam shaping to the partially coherent regime \cite{gori98,gori15}.
Applications of TGSM beams include more robust free-space optical propagation in turbulent media \cite{wang10,liu19}, sub-Rayleigh imaging \cite{Tong2012}, suppression of scintillation \cite{Wang:12} and optical trapping \cite{zhao2009}. 
{More recently, TGSM beams have been generated experimentally and their twist phase measured and controlled \cite{friberg94,wang19,tian20,wang21,canas22}. Their role as input beams in non-linear optical processes has been studied both theoretically and experimentally \cite{Walborn2020,Walborn2021,dosSantos2021,deOliveira24,Santos2025}, where it has been shown that the twist phase is transferred to the correlations of down-converted photon pairs \cite{Walborn2020,Walborn2021,dosSantos2021,Santos2025}.
\par

\par
Despite their parallel development, TGSM beams and OAM-eigenstate beams
originate from conceptually distinct frameworks. TGSM beams arise from second-order coherence theory and are characterized through their mutual coherence function. 
A central object in the description of partially coherent Gaussian beams (such as TGSMs) is the second-order moment (covariance) matrix, which can be calculated from their Wigner distribution or from their mutual intensity. For Gaussian beams such as the TGSM, this matrix completely characterizes the field: its elements encode beam widths, curvatures, coherence lengths, and correlations between spatial and angular degrees of freedom. in phase-space moment space.
\par
On the other hand, OAM eigenstates, given by
\begin{equation}
    \psi(\vec{r})=\mathcal{U}(r) e^{i\ell \phi},
    \label{eq:oameigenstate}
\end{equation}
are characterized as fully-coherent solutions of the paraxial wave equation
with well-defined OAM proportional to $\ell$. The quantum number $\ell$ is also known as the topological charge, and we use these two terms interchangeably.  In general these are not gaussian beams, and thus in principle they cannot be characterized only by considering their first and second-order moments. {For a fully coherent OAM eigenstate of the form \eqref{eq:oameigenstate}, the CM is computed from the intensity distribution $|\psi|^2$ and its phase gradients; it therefore depends on the radial profile $\mathcal{U}(r)$ through the moments $\langle r^2\rangle$ and $\langle k_r^2\rangle$.}
\par
{The connection between second-order moments and OAM has been explored in several contexts.  Gao, Wei and Weber showed that the OAM of an arbitrary beam is encoded in the cross-moments $\langle xk_y\rangle - \langle yk_x\rangle$ \cite{gaowei2000}, and Serna \emph{et al.}\ performed a complete second-order moment characterization of a doughnut-type beam, confirming agreement with LG theory \cite{serna2001}.  Simon and Mukunda's original analysis of TGSM beams identified their normal-mode decomposition in terms of LG functions \cite{simon93b,Walborn2021}, thus demonstrating that the TGSM is a statistical mixture of LG modes.
\par
The present work operates at a different level, showing that each individual coherent OAM eigenstate already has a CM with the same structure as a TGSM beam, without any decomposition. {By ``same structure'' we mean that both matrices share the same pattern of zero and nonzero entries, so that a direct term-by-term identification of beam parameters is possible. Moreover, the  $xk_y$ and $yk_x$ off-diagonal blocks relevant to angular momentum are proportional to the OAM quantum number $\ell$ and twist phase $u$, respectively.} We derive explicit parameter identifications for three canonical beam families. {Specifically,} this structural equivalence does not appear to have been stated explicitly in the literature, and provides a direct bridge between the coherent structured-beam and partially coherent communities.}
\par
{The rest of this paper is organized as follows. In Sec.~\ref{sec:TGSM} we review the TGSM covariance matrix formalism. Section~\ref{sec:CM} derives the CM of a general cylindrically symmetric OAM eigenstate and identifies it with the TGSM form, then works out the explicit parameter identifications for LG, PVB, and BG modes, derives the physical realizability conditions, examines cross-family parameter relations, and shows that the CM uniquely determines beam parameters within each family. Section~\ref{sec:propagation} discusses the consequences for beam propagation and the $M^2$ factor, showing that any TGSM beam propagation result through an ABCD system transfers directly to the matched coherent OAM beam. Section~\ref{sec:conclusion} provides concluding remarks.}

\section{Twisted Gaussian Schell Beams}
\label{sec:TGSM}
The TGSM describes gaussian beams with cross spectral density
\begin{equation}
\begin{split}
\Gamma(\mathbf{r}_1,\mathbf{r}_2)
\propto
\exp\!\Bigg[
&-\frac{(\mathbf{r}_1+\mathbf{r}_2)^2}{4\sigma^2}
-\frac{(\mathbf{r}_1-\mathbf{r}_2)^2}{2\delta^2} 
\\  & + i ku\,(\mathbf{r}_1 \times \mathbf{r}_2)_z \Bigg]
\;.
\end{split}
\end{equation}

Here $\sigma$ is the beam waist, $\delta$ is the transverse coherence length, and the real parameter $u$, known as the ``twist phase", controls the amount of
correlation-induced rotation.  
 
In a transverse plane located at longitudinal position $z$, we can define the vector of  cartesian coordinates of the transverse position ($x,y$) and wave vector ($k_x,k_y$) as $\boldsymbol{\xi}=(x,k_x,y,k_y)$. %
 In this coordinate system, the TGSM describes a gaussian beam with a  covariance matrix given by \cite{simon93,simon98}
\begin{equation}
T=\begin{pmatrix}
\sigma^2 & -\frac{k\sigma^2}{R_c} & 0 & k u\sigma^2 \\
-\frac{k\sigma^2}{R_c} &\tau^2 & -k u\sigma^2 & 0 \\
0 & -k u\sigma^2 & \sigma^2 & -\frac{k\sigma^2}{R_c} \\
k u \sigma^2 & 0 & -\frac{k\sigma^2}{R_c} & \tau^2 \\  
\end{pmatrix}, \label{eq:VTGSMdim}
\end{equation}
where $\tau^2=\frac{1}{\delta^2}+\frac{1}{4\sigma^2}+ k^2 \left( \frac{\sigma^2}{R_c^2} + u^2\sigma^2 
\right)$ is the variance of the wave vector distribution, with $R_c$ representing the radius of curvature. For the CM above to represent a bonafide TGSM beam, the twist phase must satisfy \cite{simon98}
\begin{equation}
\lvert u \rvert \leq \frac{1}{k\delta^2}.
\label{eq:ucond}
\end{equation}
By calculating $\langle L_z \rangle = \langle x k_y\rangle - \langle y k_x \rangle${, using units in which $\hbar=1$ so that $p_j = k_j$,} we see that TGSM beams carry an average orbital angular momentum (OAM) of {$2 k u \sigma^2$} per photon.

The propagation of the beams’ second moments through ABCD systems can be calculated from their covariance matrices in a single transverse plane. For simplicity, we take this plane to be $z = 0$, where the radius of curvature $R_c$ tends to infinity, causing the terms proportional to $\frac{1}{R_c}$ to vanish.   

\section{Covariance matrix of Cylindrically Symmetric Eigenmodes}
\label{sec:CM}

Let us now consider a paraxial optical beam with cylindrical symmetry, such that it be written in the form \eqref{eq:oameigenstate}. We consider beams propagating along the $z$-axis, such that $\langle x \rangle = \langle y \rangle = \langle k_x \rangle = \langle k_y \rangle = 0$ for all cases discussed here. 
Our first main result is that eigenmodes of the form \eqref{eq:oameigenstate} have covariance matrix of the form
\begin{equation}
C=\begin{pmatrix}
\frac{\langle r^2 \rangle}{2} & 0 & 0 & \frac{\ell }{2} \\
0 &\frac{\langle k_r^2\rangle}{2} & -\frac{\ell }{2} & 0 \\
0 & -\frac{\ell }{2} & \frac{\langle r^2 \rangle}{2} & 0 \\
\frac{\ell }{2} & 0 & 0 & \frac{\langle k_r^2\rangle}{2} 
\end{pmatrix}. \label{eq:CMoam}
\end{equation}
All calculations are shown in Appendix~\ref{app:genCM}.  Here 
\begin{equation}
    \langle r^2 \rangle = {2\pi} \int_0^\infty r^3 |\mathcal{U}(r)|^2 dr
\end{equation}
and 
\begin{equation}
    \langle k_r^2 \rangle = 
    {2\pi}\ \int_0^\infty \left[r\left(\frac{ d\mathcal{U}}{dr}\right)^2 + \frac{\ell^ 2}{r}\mathcal{U}^2\right] dr\;,
\end{equation} 

where in general the beam divergence depends on $\ell$ \cite{vallone206}. We see by inspection that the covariance matrix $C$ {has the same structure as} the covariance matrix $T$ of the TGSM \eqref{eq:VTGSMdim}: both have identical zero--nonzero entry patterns, which allows a direct term-by-term parameter identification without any approximation.  
\par
Comparing the elements of these two covariance matrices, we see that we can identify the beam parameters of the TGSM with those of the OAM eigenmodes.   We can associate the beam widths as
\begin{equation}
    \sigma^2 = \langle r^2 \rangle/2,
\end{equation}
\begin{equation}
    \tau^2 = \langle k_r^2 \rangle/2,
\end{equation}
the twist phase with the OAM number
\begin{equation}\label{twistphase}
    u = \frac{\ell}{2k\sigma^2}= \frac{\ell}{k \langle r^2 \rangle},
\end{equation}
and the inverse square of the coherence length as
\begin{equation}\label{colen}
    \frac{1}{\delta^2} = \tau^2 - \frac{(1 + \ell^2)}{4\sigma^2}= \frac{\langle k_r^2 \rangle}{2} - \frac{(1 + \ell^2)}{2\langle r^2 \rangle}.
\end{equation}
To make further identifications, the radial function $\mathcal{U}(r)$ must be specified.  In the following subsections, we do this for three important types of cylindrically symmetric beams. 
{%
\subsection{Second-order equivalence classes}
\label{subsec:equiv}
Equation~\eqref{eq:CMoam} establishes a universal covariance form: regardless of the
specific radial profile $\mathcal{U}(r)$, the entire second-order structure of any
cylindrically symmetric OAM eigenmode is captured by just three quantities,
$\langle r^2\rangle$, $\langle k_r^2\rangle$, and $\ell$.  This motivates defining a
second-order equivalence relation: two beams $\psi_1$ and $\psi_2$ are
\emph{second-order equivalent}, written $\psi_1\sim\psi_2$, if and only if
$C_1 = C_2$.  Under symplectic (ABCD) propagation, the covariance matrix transforms
as $C\mapsto SCS^T$, so the equivalence relation is preserved at every propagation
plane: if $\psi_1\sim\psi_2$ at $z=0$, then $\psi_1\sim\psi_2$ for all $z$.
Consequently, all second-order observables, that is the beam width, far-field divergence, $M^2$
factor, and OAM content, are identical for any two members of the same equivalence
class, independently of the detailed radial structure.  The specific beam families
considered below (LG, PVB, BG) represent distinct equivalence classes, each
parametrized by their respective values of $(\langle r^2\rangle,\langle
k_r^2\rangle,\ell)$. The TGSM beam matched to each class is identified by the
correspondence of Eqs.~\eqref{twistphase}--\eqref{colen}.}
\subsection{Laguerre-Gaussian Beams}

An important set of optical beams with OAM and of the form \eqref{eq:oameigenstate} are the
the LG modes. They are labeled by their radial and azimuthal indices $p$ and $\ell$, where the latter corresponds to the OAM value of the mode. The normalized radial function in cylindrical coordinates is given by:

\begin{equation}\label{LG}
    \mathcal{U}_{p,\ell}(r) = C_{p,\ell} \frac{2^{\frac{\ell}{2}}r^{|\ell|}}{w_0^{|\ell|+1}} L_p^{|\ell|} \left( \frac{2r^2}{w_0^2} \right) e^{-r^2 / w_0^2},
\end{equation}
where: $L_p^{|\ell|}(x)$ is the associated Laguerre polynomial of order $p$ and parameter $|\ell|$, $w_0$ is the beam waist and $C_{p,\ell} = \sqrt{\frac{2p!}{\pi(p+|\ell|)!}}$ is a normalization constant. The mode order is given by $N = 2p + |\ell|$. The Fourier Transform of an LG beam is also a Laguerre-Gaussian mode, with a beam waist given by $w_k = \frac{2}{w_0}$. 

In appendix \ref{app:LG}, we show that the CM of the LG beam is given by Eq. \eqref{eq:CMoam}, with
\begin{equation}\label{eq:r2LG}
    \langle r^2 \rangle = \frac{w_0^2}{2}\left(2p + |\ell| + 1\right) \text{, and}
\end{equation}
and
\begin{equation} \label{eq:kr2LG}
    \langle k_r^2 \rangle = \frac{2}{w_0^2}\left(2p + |\ell| + 1\right)\;.
\end{equation}
Substituting these results in eqs. \eqref{twistphase} and \eqref{colen}, we find that the twist phase and the inverse of the transverse coherence length of the corresponding TGSM beam are given by: 
\begin{equation}\label{eq:uLG}
    u = \frac{2\ell}{kw_0^2 (N+1)} \text{, and}
\end{equation}
\begin{equation}\label{eq:deltaLG}
    \frac{1}{\delta^2} = \frac{4p^2 + 4p(|\ell| + 1) + 2|\ell|}{w_0^2 (2p + |\ell| + 1)} = \frac{N(N+2) - \ell^2}{w_0^2(N+1)}\;.
\end{equation}

It now important to determine under which conditions these parameters satisfy the requirement given in Eq. \eqref{eq:ucond}, that is, to identify the necessary constraints for their CMs to correspond to those of a realizable TGSM beam. Using Eqs. \eqref{eq:uLG} and \eqref{eq:deltaLG}, we obtain the condition: 
\begin{equation} \label{eq:ucondLG}
    \frac{1}{k\delta^2} - |u| = \frac{4p^2 + 4p(|\ell| + 1)}{kw_0^2(N+1)} \ge 0\;,
\end{equation}
which is always satisfied. The maximal twist corresponds simply to $p = 0$. Therefore, the LG covariance matrix can always be reproduced by an appropriate choice of TGSM beam parameters. However, not every TGSM covariance matrix can be realized by an LG mode. To understand this, consider the following second moments product for an LG beam:
\begin{equation}\label{eq:LGconst}
    \langle x^2 \rangle \langle k_x^2 \rangle = \frac{(N+1)^2}{4}\;.
\end{equation}
Since the order $N$ is an integer, this product can assume only a discrete set of values. In contrast, the product $\sigma^2\tau^2$ can vary continuously, so an arbitrary TGSM covariance matrix will not always satisfy the constraint imposed by an LG mode.  The CM correspondence therefore does not exhaust all TGSM beams, it identifies a discrete subset matched to integer-order LG modes.

\subsection{Perfect Vortex Beams}
PVBs are optical fields whose ring radius remains almost 
independent of the topological charge $\ell$, which is valuable for applications requiring consistent mode geometry, such as OAM-multiplexed communications in structured fibers \cite{Rojas-Rojas2021,Villalba2023}.They are given by:
\begin{equation}
    \mathcal{U}_{\ell,R}(r) = A e^{-\frac{r^2 + R^2}{w^2}}I_{\ell}\left(\frac{2Rr}{w^2}\right)\;,
    \label{eq:PVB}
\end{equation}
where: $I_{\ell}(x)$ is the modified Bessel function of the first kind of order $\ell$, $R$ is the ring radius, $w$ is the ring width parameter and $A$ is a normalization constant. For $R \gg w$ and $|\ell| \ll \frac{R}{w}$, the modified Bessel function can be approximated as: $I_{\ell }\left(\frac{2Rr}{w^2}\right) \approx e^{\frac{2Rr}{w^2}}$, so that:

\begin{equation}\label{PVB}
    \mathcal{U}_{\ell,R}(r) =\mathcal{U}_R(r) = Ae^{-\frac{\left(r - R \right)^2}{w^2}}.\;
\end{equation}
This corresponds to the regime of interest, in which the intensity profile of 
the PVBs becomes effectively independent of $\ell$. Unlike LG or 
BG beams, in this regime their transverse intensity profile does not widen 
significantly as OAM increases, only the azimuthal phase changes. We note that this 
$\ell$-independence is approximate: the second-moment width of a quasi-PVB 
grows as $\sqrt{\ell}$ outside this regime \cite{vallone206}, and the ratio $R/w$ governs the 
degree to which the beam is truly ``perfect'' 
\cite{pinnell2019}.  Another difference with LG beams is that the PVB beams form a basis only in the narrow-ring limit ($w \rightarrow 0$).  This is demonstrated explicitly in appendix \ref{sec:PVBbasis}.
\par
In appendix \ref{app:PVB}, we show that, for a PVB with $R \gg w$ and $|\ell| \ll \frac{R}{w}$, we have:

\begin{equation} \label{eq:r2PVB}
    \langle r^2 \rangle = \left(R^2 + \frac{3}{4}w^2 \right) \text{, and}
\end{equation}
\begin{equation} \label{eq:kr2PVB}
    \langle k_r^2 \rangle =   \left( \frac{1}{w^2} + \frac{\ell^2}{R^2}\right)\;.
\end{equation}
Substituting these results in eqs. \eqref{twistphase} and \eqref{colen}, we find that the twist phase and the inverse of the transverse coherence length are given by: 
\begin{equation}\label{eq:uPVB}
    u = \frac{\ell}{k\left(R^2 + \frac{3}{4}w^2 \right)} \approx \frac{\ell}{kR^2} \text{, and}
\end{equation}
%\begin{align}\label{eq:deltaPVB}
    %\frac{1}{\delta^2} &= \frac{1}{2}\left( \frac{1}{w^2} + \frac{\ell^2}{R^2}\right) - \frac{(1 + \ell^2)}{2\left(R^2 + \frac{3}{4}w^2 \right)} \nonumber \\
   % &= \frac{\frac{R^2}{w^2} +\frac{3}{4}\frac{\ell^2 w^2}{R^2}-\frac{1}{4} }{2\left(R^2 + \frac{3}{4}w^2 \right)}  \nonumber \\
    %&= \frac{1 + \frac{3\ell^2 w^4}{4R^4} - \frac{w^2}{4R^2}}{2w^2\left(1 + \frac{3w^2}{4R^2}\right)} \approx \frac{1}{2w^2}\;.
%\end{align}

\begin{equation}\label{eq:deltaPVB}
    \frac{1}{\delta^2} = \frac{1}{2}\left( \frac{1}{w^2} + \frac{\ell^2}{R^2}\right) - \frac{(1 + \ell^2)}{2\left(R^2 + \frac{3}{4}w^2 \right)} \approx \frac{1}{2w^2}\;.
\end{equation}

\par In the case of PVBs, using Eqs. \eqref{eq:uPVB} and \eqref{eq:deltaPVB}, the TGSM realizability condition given in Eq. \eqref{eq:ucond} leads to the inequality: 
\begin{equation}
    2|\ell| \leq \frac{R^2}{w^2}\;, 
\end{equation}
which is automatically satisfied in domain of validity of the approximations used to compute the PVB second moments \cite{pinnell2019}. The maximal twist case lies outside of this domain. 

Since $\frac{R^2}{w^2} \gg 1$, it follows that $\langle x^2 \rangle \langle k_x^2 \rangle \gg 1$. Imposing this condition on the TGSM beams parameters leads to $\sigma^2 \tau^2 \gg 1$. Because the twist parameter $u$ is limited by the requirement given in \eqref{eq:ucond}, the dominant contribution to $\tau^2$ comes from the term $\frac{1}{\delta^2}$. Consequently,
\begin{equation}\label{eq:sigmadelta}
    \frac{\sigma^2}{\delta^2} \gg 1\;,
\end{equation}
meaning that {to describe a PVB via the covariance matrix formalism in this regime, one should use the parameter identifications given in Table~\ref{table}, which correspond to a low-coherence TGSM. The matched TGSM for a PVB is therefore necessarily in the low-coherence regime.}

\subsection{Bessel-Gauss Beams}
BG beams are optical fields formed by modulating an ideal (non-diffracting) Bessel beam with a Gaussian envelope, yielding a physically realizable beam that maintains a narrow central core and exhibits extended propagation invariance. The combination of their quasi-non-diffracting propagation, self-healing behavior and well-defined orbital angular momentum makes these beams attractive for free-space optical communication systems \cite{ahmed2016}.

They are of the form \eqref{eq:oameigenstate}, with a radial function given by:
\begin{equation}\label{BG}
    \mathcal{U}_\ell(r) = B_\ell J_\ell\left(rk_b\right) e^{-\frac{r^2}{w_b^2}}\;,
\end{equation}
where: $J_\ell(x)$ is the Bessel function of the first kind of order $\ell$, $B_\ell$ is a normalization constant and $w_b$ is the Gaussian beam waist. In appendix \ref{app:BG}, we show that, for a BG beam with $w_b \gg \frac{1}{k_b}$, and  $|\ell| \ll \frac{w_bk_b}{2}$, corresponding to the regime in which many oscillations of the Bessel function occur within the gaussian envelope, we have:
\begin{equation}\label{eq:r2BG}
    \langle r^2 \rangle = \left(\frac{w_b^2}{4} + \frac{\ell^2}{k_b^2} \right), 
\end{equation}
and
\begin{equation}\label{eq:kr2BG}
\langle k_r^2 \rangle = \left(k_b^2 + \frac{3}{w_b^2} \right)\;.
\end{equation}
%\spw{Check: the BG--PVB duality $\langle r^2\rangle_\text{PVB}\leftrightarrow\langle k_r^2\rangle_\text{BG}$ under $R\to k_b$, $w\to 2/w_b$ applied to $\langle r^2\rangle_\text{PVB} = R^2 + \frac{3w^2}{4}$ gives $\langle k_r^2\rangle_\text{BG} = k_b^2 + \frac{3}{w_b^2}$, \emph{not} $k_b^2 + \frac{3}{2w_b^2}$. The appendix result (Eq.~\eqref{eq:kx2BG}) and the result stated here appear to be inconsistent by a factor of 2 in the second term. One of these two equations needs to be corrected. Please verify independently.}
Substituting these results in eqs. \eqref{twistphase} and \eqref{colen}, we find that the twist phase and the inverse of the transverse coherence length are given by: 
\begin{equation}\label{eq:uBG}
    u = \frac{\ell}{k\left(\frac{1}{4}w_b^2 + \frac{\ell^2}{k_b^2} \right)} \approx \frac{4 \ell}{kw_b^2}, 
\end{equation}
and
\begin{equation}\label{eq:deltaBG}
    \frac{1}{\delta^2} = \frac{1}{2}\left(k_b^2 + \frac{3}{w_b^2} \right) - \frac{(1 + \ell^2)}{2\left( \frac{1}{4}w_b^2 + \frac{\ell^2}{k_b^2} \right)} \approx \frac{k_b^2}{2} \;.
\end{equation}
%\spw{This expression for $1/\delta^2$ uses $\langle k_r^2\rangle_\text{BG}$ from Eq.~\eqref{eq:kr2BG}. If that equation has a factor-of-2 error in the $3/(2w_b^2)$ term, the full expression here changes, though the leading-order approximation $\approx k_b^2/2$ is unaffected since it drops the $3/w_b^2$-type term entirely.}

%\begin{align}
    %\frac{1}{\delta^2} &= \frac{1}{2}\left(k_b^2 + \frac{3}{2w_b^2} \right) - \frac{(1 + \ell^2)}{2\left( \frac{1}{4}w_b^2 + \frac{\ell^2}{k_b^2} \right)}\nonumber \\ 
    %&= \frac{\frac{1}{4}w_b^2 k_b^2 + \frac{3\ell^2}{2w_b^2k_b^2}-\frac{5}{8}}{2\left(\frac{1}{4}w_b^2 + \frac{\ell^2}{k_b^2} \right)}\nonumber \\
    %&= \frac{k_b^2 \left(\frac{1}{4} + \frac{3\ell^2}{2w_b^4k_b^4} - \frac{5}{8w_b^2k_b^2} \right)}{2\left( \frac{1}{4} + \frac{\ell^2}{k_b^2 w_b^2} \right)} \;.
%\end{align}

Substituting Eqs. \eqref{eq:deltaBG} and \eqref{eq:uBG} into the condition \eqref{eq:ucond} gives:
\begin{equation}
    8|\ell| \leq w_b^2 k_b^2\;.
\end{equation}

As in the case of perfect vortex beams, the inequality is automatically satisfied within the domain of validity of the approximations, and the maximal twist scenario is outside of this domain. Similarly to the PVBs, the condition in \eqref{eq:sigmadelta} must be fulfilled. This means that to describe a BG beam via the covariance matrix formalism in the regime of interest, one should employ the parameter identifications given in Table~\ref{table}, which correspond to a low-coherence TGSM beam. In other words, within this regime, the matched TGSM for a BG beam or PVB necessarily exhibits low coherence, and the full TGSM propagation toolbox applies under that identification. 

\subsection{Comparison and summary}
The three families represent three qualitatively different regimes of the
coherence--OAM relationship, with parameter correspondence summarized in table~\ref{table}.  For LG modes, $\delta$ and $u$ are coupled
through the mode indices. Changing $\ell$ at fixed $p$ changes both simultaneously,
and only at $p=0$ is the coherence length infinite.
For PVBs and BG beams, $\delta$ is decoupled from $\ell$ in the leading
approximation, such that the OAM and the effective coherence length are independent
degrees of freedom of the matching TGSM.
This decoupling is intimately connected to the condition $\sigma^2/\delta^2\gg 1$
(Eq.~\eqref{eq:sigmadelta}), satisfied by both PVBs and BG beams in the regime of
interest. When the TGSM is highly incoherent relative to its beam size, the
dominant contribution to $\tau^2$ comes from $1/\delta^2$ rather than the twist
term $k^2u^2\sigma^2$, so spatial structure and OAM content genuinely separate.

%%%%%%%%%%%%%%%

\subsection{Parameter Determination from the Covariance Matrix Within a Mode Family}

Although a single covariance matrix may represent both a TGSM beam and a coherent OAM beam, as shown above, it uniquely determines the TGSM beam through the parameters $\sigma$, $u$, and $\delta$ \cite{simon93}. Thus, two distinct TGSM beams cannot share the same CM. The LG, PVB and BG beams are non-gaussian functions, in general, and thus it is to be expected that they cannot be determined by the elements of the CM alone.  However, somewhat surprisingly, we note here that the CM is indeed sufficient in some cases.  First, for all beams of the form \eqref{eq:oameigenstate}, the parameter $\ell$ follows directly from the anti-diagonal elements of Eq. \eqref{eq:CMoam}, so that only the remaining parameters need to be assessed: $w_0$ and $p$ for LG modes, $w$ and $R$ for PVBs, and $w_b$ and $k_b$ for BG beams.

For an LG beam, the radial parameter $p$ can be determined from the product $\langle x^2 \rangle \langle k_x^2 \rangle$, as shown in Eq. \eqref{eq:LGconst}. Since $\ell$ is already fixed and $p \geq 0$, this relation specifies $p$ unambiguously. Substituting this value into Eq. \eqref{eq:r2LG} or Eq. \eqref{eq:kr2LG} then determines $w_0$. Therefore, the covariance matrix uniquely determines the LG mode parameters.

For both PVB and BG beams, the diagonal elements of the covariance matrix lead to a nonlinear system relating the beam parameters, which may admit multiple solutions. In the regime of interest, however, the dominant terms yield relations in which each parameter is approximately determined by a single second moment. For PVBs, Eqs. \eqref{eq:r2PVB} and \eqref{eq:kr2PVB} give $\langle r^2 \rangle \approx R^2$ and $\langle k_r^2 \rangle \approx \frac{1}{w^2}$. Similarly, for BG beams, Eqs. \eqref{eq:r2BG} and \eqref{eq:kr2BG} yield $\langle r^2 \rangle \approx \frac{w_b^2}{4}$ and $\langle k_r^2 \rangle \approx k_b^2$. Thus, within this approximation, the covariance matrix determines the beam parameters to leading order for both families. A rigorous proof of exact uniqueness from the full, unapproximated moment equations is beyond the scope of this work. The leading-order inversions above are sufficient for practical parameter estimation in the regimes considered.

}

%%%%%%%%%%%%%%%%%

\subsection{Cross-family parameter correspondences}

Table \ref{table} summarizes the correspondence between the covariance matrix parameters of the TGSM beam and those of the cylindrically symmetric beams analyzed in this work. The same comparison of matrix elements, when applied to the coherent beam families, provides the parameter relations under which their CMs coincide. 
A comparison of the LG mode and PVB covariance matrices in the regime $R \gg w$ and $|\ell| \ll \frac{R}{w}$ leads to:
\begin{align}
    w &\approx \frac{w_0}{\sqrt{2(N+1)}}\;,  \\
    R &\approx \frac{w_0\sqrt{N+1}}{\sqrt{2}}\;.
\end{align}

Similarly, equating the LG and BG beams covariance matrices in the regime $w_b \gg \frac{1}{k_b}$ and $|\ell| \ll \frac{w_bk_b}{2}$ results in
\begin{align}
    w_b &\approx w_0\sqrt{2(N+1)}\;, \label{eq:LGBG1} \\
    k_b &\approx \frac{\sqrt{2(N+1)}}{w_0}\;.  \label{eq:LGBG2}
\end{align}
Using the relations derived above, we obtain $\frac{R}{w} \sim N + 1$ and $w_bk_b \sim 2(N+1)$. Therefore, the approximations employed for the PVB and BG beams are translated to $N \gg 1$ for the LG mode. 

Finally, comparing the covariance matrix elements of the PVBs and BG beams  under the approximations stated above yields
\begin{align}
    R &\approx \frac{w_b}{2}\;, \\
    \frac{1}{w} &\approx k_b\;.
\end{align}

\renewcommand{\arraystretch}{2.8}

\begin{table*}
\centering
\begin{tabular}{|c|c|c|c|}
\hline
\textbf{TGSM} & \textbf{LG} & \textbf{PVB} & \textbf{BG} \\ \hline

$\sigma^2$
&
$\displaystyle \frac{w_0^2}{4}\left(N + 1\right)$
&
$\displaystyle \frac{1}{2}\left(R^2 + \frac{3}{4}w^2 \right)$
&
$\displaystyle \frac{1}{2}\left(\frac{w_b^2}{4} + \frac{\ell^2}{k_b^2} \right)$
\\ \hline

$\tau^2$
&
$\displaystyle \frac{1}{w_0^2}\left(N + 1\right)$
&
$\displaystyle \frac{1}{2}\left( \frac{1}{w^2} + \frac{\ell^2}{R^2} \right)$
&
$\displaystyle \frac{1}{2}\left(k_b^2 + \frac{3}{w_b^2} \right)$
\\ \hline

$u$
&
$\displaystyle \frac{2 \ell}{k w_0^2 (N+1)}$
&
$\displaystyle \frac{\ell}{kR^2}$
&
$\displaystyle \frac{4\ell}{kw_b^2}$
\\ \hline

$\displaystyle \frac{1}{\delta^2}$
&
$\displaystyle \frac{N(N+2)-\ell^2}{w_0^2(N+1)}$
&
$\displaystyle \frac{1}{2w^2}$
&
$\displaystyle \frac{k_b^2}{2}$
\\ \hline

\end{tabular}

\caption{Correspondence between covariance matrix parameters for TGSM beams, LG beams, Perfect Vortex beams and Bessel-Gauss beams.}

\label{table}

\end{table*}

{\section{Beam propagation and $M^2$ factor equivalence}
\label{sec:propagation}

A direct consequence of the CM equivalence is that the full TGSM propagation toolbox applies to any matched coherent OAM beam. Once the parameter identifications of Sec.~\ref{sec:CM} are made, all TGSM results for beam-width evolution, Rayleigh range, and far-field divergence under arbitrary ABCD systems carry over without further calculation. For a beam propagating in free space, the covariance matrix
evolves as $C(z) = S(z)\,C(0)\,S(z)^T$, where $S(z)$ is the symplectic transfer matrix
for free-space propagation a distance $z$ \cite{DENG2004205}.
 Because two beams with the same $C(0)$ are mapped by the same $S(z)$, they have the same $C(z)$ at every plane. In particular, they share
the same beam-width evolution, the same Rayleigh range, and the same far-field divergence angle.

\par
At the reference plane $z = 0$, chosen such that the position--momentum correlation vanishes, a compact measure of this propagation equivalence is the beam-quality parameter
\begin{equation}\label{eq:M2defISO}
  M^2 = \frac{\pi \sigma_w(0)\theta_\infty}{\lambda}\,,
\end{equation}
where $\sigma_w(z) = \sqrt{2\langle r^2\rangle (z)}$ is the second-moment beam radius and $\theta_\infty = \sqrt{2\langle k_r^2 \rangle (0)/k^2}$ is the far-field divergence half-angle, with both definitions holding for cylindrically symmetric beams. Substituting these expressions into Eq. \eqref{eq:M2defISO} and using $k = \frac{2\pi}{\lambda}$, we obtain
\begin{equation}\label{eq:M2def}
  M^2 = \sqrt{\langle r^2\rangle (0)\,\langle k_r^2\rangle (0)}\,,
\end{equation}
which is invariant under symplectic transformations.}
This definition follows the second-moment convention of Refs.~\cite{simon93,simon98}, where $M^2=1$ for a fundamental Gaussian beam.

\section{Conclusion}
\label{sec:conclusion}

In this work we established a correspondence between the covariance matrices of twisted Gaussian--Schell model beams and those of coherent OAM-carrying beams, including Laguerre--Gaussian, Bessel--Gaussian, and perfect vortex beams, showing that these matrices share the same structural form. As a consequence, the covariance matrix alone may not uniquely identify the family to which an OAM beam belongs. {More broadly, Eq.~\eqref{eq:CMoam} defines a second-order equivalence relation
on cylindrically symmetric OAM beams: two such beams are second-order equivalent if
and only if they share the same values of $\langle r^2\rangle$, $\langle
k_r^2\rangle$, and $\ell$, and this equivalence is preserved under all symplectic
propagation.  The coherent beam families studied here and the matched TGSM beams
belong to the same equivalence classes in this sense.} In particular, within the regimes considered here, a single covariance matrix can represent Bessel--Gaussian and perfect vortex beams, as well as high-order Laguerre--Gaussian modes. We also examined the conditions under which the same covariance matrix can describe both a TGSM beam and a coherent OAM beam. {Because two beams sharing a covariance matrix are mapped identically by any symplectic (ABCD) transformation, they exhibit the same beam-width evolution, Rayleigh range, far-field divergence, and $M^2$ beam-quality factor.} This expands the experimental flexibility in situations where only the beam's orbital angular momentum and spreading are relevant, and the detailed radial structure of the field is not important. {Conversely, the correspondence provides a prescription for designing a partially coherent TGSM beam that mimics the propagation and OAM content of any specified coherent OAM eigenstate, without reproducing its full field structure.}
We further examined the invertibility of the covariance matrix within each beam family. Surprisingly, for LG modes, the beam parameters are uniquely and exactly determined by the second-order moments. For PVB and BG beams, the CM determines the beam parameters to leading order in the relevant approximation regime, though a proof of exact uniqueness from the full moment equations remains an open question.
{We also showed in the appendix that the PVB modes form a complete orthonormal basis in the narrow-ring
limit ($w \to 0$).  The practical condition
for the PVB expansion to faithfully represent a given field is $w \ll\lambda_{\min}$,
where $\lambda_{\min}$ is the shortest transverse length scale of the field,
with the OAM spectrum being reproduced exactly for any $w$.} In addition to making a fundamental connection between partially coherent TGSM beams and coherent OAM eigenmodes, we expect our results to be useful in structured-light applications. 

%%%%%%%%%

\section{Appendix: Calculations for Cylindrically Symmetric Beams}
\label{app:genCM}

Let us consider a cylindrically symmetric optical beam described by the wavefunction:
\begin{equation}\label{eq:generalbeam}
    \psi(r, \phi) = \mathcal{U}(r) e^{i \ell\phi}\;.
\end{equation}
For simplicity, we assume $\mathcal{U}(r)$ is real, as is the case at $z = 0$ for the LG modes, BG beams and PVBs. The following results remain valid under the multiplication of the radial profile by a constant phase factor. 

For $\psi(r,\phi)$ to be normalized, we must have
\begin{equation}\label{eq:Norm}
    \int_0^\infty \left|\mathcal{U}(r) \right|^2 r\,dr = \frac{1}{2\pi}\;.
\end{equation}

The transverse momentum operators in the $x$ and $y$ directions, expressed in polar coordinates ($x = r\cos\phi \,, \; y = r\sin\phi$), are 
\begin{align}
    k_x = -i  \frac{\partial}{\partial x} = -i\left(\cos\phi \,\frac{\partial}{\partial r} - \frac{\sin\phi}{r}\, \frac{\partial}{\partial \phi}\right) \label{eq:delx} \;,\\
    k_y = -i  \frac{\partial}{\partial y} = -i\left(  \sin\phi \,\frac{\partial}{\partial r} + \frac{\cos\phi}{r} \,\frac{\partial}{\partial \phi} \right) \;.\label{eq:dely}
\end{align}
Evaluating the derivatives of Eq. \eqref{eq:generalbeam} using Eqs. \eqref{eq:delx} and \eqref{eq:dely}, we obtain
\begin{align}
    \frac{\partial \, \psi}{\partial x} &= e^{i\ell \phi}\left[ \cos\phi \, \frac{d\,\mathcal{U}}{d r} - i\ell\frac{\sin\phi}{r}\,\mathcal{U} \right] , \label{eq:delxPsi}\\
    \frac{\partial \, \psi}{\partial y} &= e^{i\ell \phi}\left[ \sin\phi \,\frac{d\,\mathcal{U}}{d r} + i\ell\frac{\cos\phi}{r}\,\mathcal{U} \right]. \label{eq:delyPsi}
\end{align}
The second derivatives are
\begin{align}
\begin{split}
    \frac{\partial^2 \psi}{\partial x^2} = e^{i\ell \phi} &\Bigg[ \cos^2\phi \,\frac{d^2\mathcal{U}}{d r^2} + \sin^2\phi\left(\frac{1}{r}\,\frac{d\,\mathcal{U}}{d r} - \frac{\ell^2}{r^2}\,\mathcal{U} \right) \\ &-2i\ell\,\sin\phi\cos\phi \left(\frac{1}{r}\,\frac{d\,\mathcal{U}}{d r} - \frac{1}{r^2}\mathcal{U} \right)    \Bigg]\;,\label{eq:del2xPsi} 
\end{split}
  \\
\begin{split}
\frac{\partial^2 \psi}{\partial y^2} = e^{i\ell \phi} &\Bigg[ \sin^2\phi \,\frac{d^2\mathcal{U}}{d r^2} + \cos^2\phi\left(\frac{1}{r}\,\frac{d\,\mathcal{U}}{d r} - \frac{\ell^2}{r^2}\,\mathcal{U} \right) \\ &+2i\ell\,\sin\phi\cos\phi \left(\frac{1}{r}\,\frac{d\,\mathcal{U}}{d r} - \frac{1}{r^2}\mathcal{U} \right) 
    \Bigg] \;,\label{eq:del2yPsi}
\end{split} \\
\begin{split}
    \frac{\partial^2\psi}{\partial y \,\partial x} = e^{i\ell \phi} &\Bigg[\sin\phi\cos\phi \left( \frac{d^2\mathcal{U}}{d r^2} - \frac{1}{r}\,\frac{d\,\mathcal{U}}{d r} + \frac{\ell^2}{r^2}\mathcal{U}     \right) \\ &- i\ell (\sin^2\phi - \cos^2\phi) \left( \frac{1}{r}\,\frac{d\,\mathcal{U}}{d r} - \frac{1}{r^2}\,\mathcal{U} \right)
    \Bigg] \;. \label{eq:delxyPsi}
\end{split}
\end{align}

\subsection{Expectation Value of $x k_x $}

Using the derivative \eqref{eq:delxPsi}, the expectation value
\begin{equation}
\begin{split}
    \langle xk_x \rangle   =& -i \int_0^\infty r^2 \, \mathcal{U}(r)\, \Bigg[ \int_0^{2\pi} \bigg( \cos^2\phi \frac{d\, \mathcal{U}(r)}{d r} \\ &+\frac{i\ell}{r} \sin\phi\cos\phi \, \mathcal{U}(r) \bigg) d\phi \Bigg] dr\;.
\end{split}
\end{equation}
Using well known integrals of trigonometric functions, the integral becomes
\begin{equation}
    \langle x k_x \rangle = -i  \pi \int_0^\infty r^2 \: \mathcal{U}(r) \frac{d\, \mathcal{U}(r)}{d r}\, dr\;,
\end{equation}
This gives:
\begin{equation}
    \langle x k_x \rangle = -\frac{i \pi}{2} \Biggl( \big[ r^2 \, \left|\mathcal{U}(r)\right|^2 \big]_0^\infty - 2\int_0^\infty r \, \mathcal{U}^2 (r) dr \Biggr) \;.
\end{equation}
For physically admissible beams, $\mathcal{U}(r)$ does not diverge for $r = 0$ and decays sufficiently fast as $r \rightarrow\infty$, so that the boundary term vanishes. Using the normalization condition in Eq. \eqref{eq:Norm}, we obtain:
\begin{equation}\label{xk_x}
    \langle x k_x \rangle = \frac{i}{2}\;.
\end{equation}

Since $xk_x$ is not Hermitian, the physically meaningful second moment is defined using the symmetrized operator $\frac{1}{2}(xk_x + k_x x)$. Using the commutation relation $[x,k_x] = i$, we see that its expectation value vanishes.  

\subsection{Expectation Value of $x k_y $}

Using equation \eqref{eq:delyPsi}, we have, in polar coordinates
\begin{equation}
\begin{split}
\langle x k_y \rangle &= -i \int_0^\infty \int_0^{2\pi} \mathcal{U}(r) \, r^2  \sin\phi\cos\phi \,\frac{\partial \mathcal{U}(r)}{\partial r} \,d\phi \,dr \\
&- i \int_0^\infty \int_0^{2\pi} |\mathcal{U}(r)|^2 \,r \,\,(i\ell)\cos^2\phi  \,d\phi \,dr.
\end{split}
\end{equation}
Using well-known integrals of trigonometric functions, the first term vanished and the remaining contribution reduces to
\begin{equation}
\langle x k_y \rangle = -i (i\ell)\, \pi \int_0^\infty |\mathcal{U}(r)|^2 \,r \,dr.
\end{equation}
Finally, since the function is normalized, as seen in Eq. \eqref{eq:Norm}, we obtain:
\begin{equation}\label{xk_y}
\langle x k_y \rangle = \frac{\ell}{2}\;.
\end{equation}

The calculation for $\langle y k_x \rangle$ is analogous and yields
\begin{equation}\label{yk_x}
\langle y k_x \rangle = -\frac{\ell}{2}\;.
\end{equation}

\subsection{Expectation Value of $x^2$}
Using the wavefunction equation \eqref{eq:generalbeam} and working in polar coordinates, we obtain:
\begin{equation}
    \langle x^2 \rangle = \int_0^\infty \int_0^{2\pi} r^2 \cos^2\phi \,|\mathcal{U}(r)|^2\, r \, d\phi \, dr\;.
\end{equation}
After integration in $\phi$ we have
\begin{equation}\label{x2}
    \langle x^2 \rangle = \pi \int_0^\infty r^3 |\mathcal{U}(r)|^2 \,dr\;.
\end{equation}
By symmetry, $\langle x^2 \rangle = \langle y^2 \rangle=\langle r^2 \rangle/2$, where we define:
\begin{equation}
    \langle r^2 \rangle = 2\pi \int_0^\infty r^3 |\mathcal{U}(r)|^2 \,dr\;.
\end{equation}

\subsection{Expectation Value of $xy$}
In polar coordinates, the expectation value of $xy$ can be written as:
\begin{equation}
    \langle xy \rangle =\int_0^\infty \int_0^{2\pi} r^3 \sin\phi \cos\phi \,|\mathcal{U}(r)|^2 \,d\phi\, dr\;.
\end{equation}
The angular integral vanishes, resulting in
\begin{equation}\label{xy}
    \langle xy \rangle = 0\;.
\end{equation}

\subsection{Expectation Value of $k_x^2$}

Using the derivative in Eq. \eqref{eq:del2xPsi}, we obtain
\begin{equation}
    \begin{split}
    \langle k_x^2 \rangle = &- \int_0^\infty \int_0^{2\pi} r\,\mathcal{U}(r)\, \Bigg[ \cos^2\phi \,\frac{d^2\mathcal{U}(r)}{d r^2} \\&+ \sin^2\phi\left(\frac{1}{r}\,\frac{d\,\mathcal{U}(r)}{d r} - \frac{\ell^2}{r^2}\,\mathcal{U}(r) \right) \\ 
    &-2\,i\ell\,\sin\phi\cos\phi \left(\frac{1}{r}\,\frac{d\,\mathcal{U}(r)}{d r} - \frac{1}{r^2}\mathcal{U}(r) \right)  \Bigg] \,dr\,d\phi\;.
    \end{split}
\end{equation}
Noting that the last term vanishes and performing straightforward integration, this reduces to
\begin{equation}\label{eq:k_xk_xOAM}
    \langle k_x^2 \rangle = \pi\ \int_0^\infty \left[r\left(\frac{ d\,\mathcal{U}(r)}{dr}\right)^2 + \frac{\ell^ 2}{r}\, \mathcal{U}^2(r) \right] dr\;.
\end{equation}

By symmetry, we have $\langle k_x^2 \rangle = \langle k_y^2 \rangle =  \langle k_r^2 \rangle/2$, where we define:
\begin{equation}
    \langle k_r^2 \rangle = 2\pi\ \int_0^\infty \left[r\left(\frac{ d\,\mathcal{U}(r)}{dr}\right)^2 + \frac{\ell^ 2}{r}\, \mathcal{U}^2(r) \right] dr\;.
\end{equation}

\subsection{Expectation Value of $k_yk_x$}

Using the derivative from Eq. \eqref{eq:delxyPsi}, the expectation value of $k_yk_x$ reads:
\begin{equation}
    \begin{split}
    &\langle k_y k_x \rangle = -\int_0^{\infty}\int_0^{2\pi} r\,\mathcal{U}(r) \\&\Biggr[\sin\phi\cos\phi \left( \frac{d^2\mathcal{U}(r)}{d r^2} - \frac{1}{r}\,\frac{d\,\mathcal{U}(r)}{d r} + \frac{\ell^2}{r^2}\mathcal{U}(r)     \right) \\ 
    &+ i\ell (\sin^2\phi - \cos^2\phi) \left( \frac{1}{r}\,\frac{d\,\mathcal{U}(r)}{d r} - \frac{1}{r^2}\,\mathcal{U}(r) \right)
    \Biggr] \,d\phi \,dr\;.
    \end{split}
\end{equation}
The angular integrals vanish, so $\langle k_y k_x \rangle = 0$.

\subsection{Fourier Transform}

The two-dimensional Fourier transform of a function $f(x,y)$ in cartesian coordinates is given by:

\begin{equation}
F(k_x, k_y) = \frac{1}{2\pi}\int \int f(x, y) e^{-i (k_x x + k_y y)} \,dx \,dy.
\end{equation}

In polar coordinates, with:
\begin{align}
x &= r \cos\phi, \quad y = r \sin\phi, \notag\\
k_x &= k_r \cos\theta, \quad k_y = k_r \sin\theta,
\end{align}
the Fourier transform of a function of the form given in Eq. \eqref{eq:generalbeam} becomes:
\begin{equation}
F(k_r, \theta) = \frac{1}{2\pi}\int_0^{\infty} \int_0^{2\pi} \mathcal{U}(r) e^{i \ell \phi} e^{-i k_r r \cos(\phi - \theta)} r \, d\phi \, dr.
\end{equation}

Using the Jacobi--Anger expansion:
\begin{equation}
e^{-i k_r r \cos(\phi - \theta)} = \sum_{m=-\infty}^{\infty} (-i)^m J_m(k_r r) e^{i m (\phi - \theta)},
\end{equation}
where $J_m(k_r r)$ is the Bessel function of the first kind of order $m$, we can write the angular integral as:
\begin{equation}
I = \int_0^{2\pi} e^{i \ell \phi} \sum_{m=-\infty}^{\infty} (-i)^m J_m(k_r r) e^{i m (\phi - \theta)} d\phi.
\end{equation}
Interchanging summation and integration, we have
\begin{equation}
I = \sum_{m=-\infty}^{\infty} (-i)^m J_m(k_r r) \int_0^{2\pi} e^{i (\ell + m) \phi} e^{-i m \theta} d\phi.
\end{equation}
Using the orthogonality condition:
\begin{equation}
\int_0^{2\pi} e^{i (\ell + m) \phi} d\phi = 2\pi \delta_{\ell, -m},
\end{equation}
this simplifies to:
\begin{equation}
I = 2\pi (-i)^{-\ell} J_{-\ell}(k_r r) e^{i \ell \theta}.
\end{equation}

Noting $J_{-n}(x) = (-1)^n J_n(x)$ for $n \in \mathds{Z}$, we obtain:
\begin{equation}
I = 2\pi (-i)^{\ell} J_{\ell}(k_r r) e^{i \ell \theta}.
\end{equation}
Substituting this result back into the Fourier transform gives
\begin{equation}
F(k_r, \theta) = \int_0^{\infty} \mathcal{U}(r) \cdot (-i)^{\ell} J_\ell(k_r r) e^{i \ell \theta} r \, dr\;.
\end{equation}

Defining the Hankel transform of order $\ell$:
\begin{equation}\label{eq:Hankel}
\mathcal{H}_\ell(k_r) = \int_0^{\infty} \mathcal{U}(r) J_\ell(k_r r)\, r \, dr\;,
\end{equation}
we arrive at the final result:
\begin{equation}\label{FT}
F(k_r, \theta) = (-i)^{\ell} e^{i \ell \theta} U_\ell(k_r).
\end{equation}
This shows that the Fourier transform preserves the azimuthal phase structure $e^{i \ell \theta}$ and the radial dependence follows a Hankel transform.

%%%%%%%%%%%%
\section{Appendix: Laguerre-Gaussian Beams} \label{app:LG}

The normalized LG mode at $z = 0$ is given by \eqref{LG}. 

\subsection{Fourier Transform of an LG Beam}
Substituting the LG mode $\mathcal{U}_{p,\ell}$ into \eqref{eq:Hankel} gives:
\begin{equation}\label{eq:HankLG}
    \mathcal{H}_{\ell} (k_r) = C \frac{2^{\frac{|\ell|}{2}}}{w_0^{|\ell|+1}} \int_0^{\infty} r^{|\ell|+1}J_{\ell}(k_rr) \,L_p^{|\ell|}\left(\frac{2r^2}{w_0^2}\right)e^{-\frac{r^2}{w_0^2}} \;.
\end{equation}
For $\ell < 0$, the relation $J_{-m}(x) = (-1)^mJ_m(x)$ can be used to express the integral in terms of $|\ell|$ only. Applying identity 7.421.4 of Ref. \cite{GR:2007} then yields
\begin{equation}
\begin{split}
     \mathcal{H}_{\ell} (k_r) &= C \, 2^\frac{|\ell|}{2}\left(\frac{w_0}{2}\right)^{|\ell|+1}k_r^{|\ell|}  \,L_p^{|\ell|}\left(\frac{k_r^2w_0^2}{2}\right)e^{-\frac{k_r^2w_0^2}{4}}\;, \\ &(\ell > 0)\;,\\
     \mathcal{H}_{\ell} (k_r) &= (-1)^{|\ell|} C\,  2^\frac{|\ell|}{2}\left(\frac{w_0}{2}\right)^{|\ell|+1}k_r^{|\ell|}  \,L_p^{|\ell|}\left(\frac{k_r^2w_0^2}{2}\right)e^{-\frac{k_r^2w_0^2}{4}}\;,\\ &(\ell < 0)\;.
\end{split}
\end{equation}
Substituting this result into Eq. \eqref{FT} and noting that
\begin{equation}
   (-1)^{|\ell|}(-i)^{-|\ell|} = (-1)^{|\ell|}i^{|\ell|}  = (-i)^{|\ell|} \;,
\end{equation}
the Fourier transform of an LG mode becomes
\begin{equation}\label{LGFT}
\begin{split}
    F_{\ell,p}(k_r,\theta) = &(-i)^{|\ell|}   C\, 2^\frac{|\ell|}{2}\left(\frac{w_0}{2}\right)^{|\ell|+1}k_r^{|\ell|} \\ &\times L_p^{|\ell|}\left(\frac{k_r^2w_0^2}{2}\right)e^{-\frac{k_r^2w_0^2}{4}}e^{i\ell\theta}.
\end{split}
\end{equation}
This shows that the Fourier Transform of an LG beam is itself a Laguerre-Gaussian mode with the same indices $p$ and $\ell$, and with a beam waist given by $w_k = \frac{2}{w_0}$.
%\begin{equation}\label{eq:LGFT2}
    %F_{\ell,p}(k_r,\theta) = (-i)^{|\ell|}   C\,\frac{ 2^\frac{|\ell|}{2}}{w_k^{|\ell|+1}}k_r^{|\ell|} L_p^{|\ell|}\left(\frac{2k_r^2}{w_k^2}\right)e^{-\frac{k_r^2}{w_k^2}}e^{-i\ell\theta}.
%\end{equation}

%%%%%%%%%%%

\subsection{Expectation Value of $x^2$}

Substituting the LG mode $\mathcal{U}_{p,\ell}$ into Eq. \eqref{x2}, we obtain:
\begin{equation}
\begin{split}
    \langle x^2 \rangle = \frac{C^2\, 2^{|\ell|}\pi}{w_0^{2|\ell|+2}}\int_0^\infty r^{2|\ell|+3}\left[L_p^{|\ell|}\left(\frac{2r^2}{w_0^2}\right) \right]^2e^{-\frac{2r^2}{w_0^2}}\,dr\;.
\end{split}
\end{equation}
Introducing the change of variable $a = \frac{2r^2}{w_0^2}\rightarrow da = \frac{4r}{w_0^2}\,dr$, the integral becomes:
\begin{equation}
    \langle x^2 \rangle = \frac{C^2\, w_0^2\pi}{8}\int_0^\infty a^{|\ell|+1}\left[L_p^{|\ell|}(a)\right]^2 e^{-a}\,da\;.
\end{equation}
The integral above can be evaluated using the known identity for associated Laguerre polynomials:
\begin{equation}
\begin{split}
    \int_0^\infty& \: a^{|\ell|+1}\left[L_p^{|\ell|}(a)\right]^2e^{-a}\,da = \\ &\frac{(p+|\ell|)!}{p!}(2p+|\ell|+1)\;.
\end{split}
\end{equation}
Substituting the normalization constant $C^2 = \frac{2p!}{\pi(p+|\ell|)!}$, we finally obtain:
\begin{equation}\label{eq:x2LG}
    \langle x^2 \rangle = \frac{w_0^2}{4}(2p + |\ell| +1)\;.
\end{equation}

\subsection{Expectation Value of $k_x^2$}

Considering that the integral defining $\langle k_x^2\rangle$ in the Fourier space is analogous to that of $\langle x^2\rangle$ in the real space, from Eqs. \eqref{LGFT} and \eqref{eq:x2LG}, we find:
\begin{equation}\label{eq:px2LG}
    \langle k_x^2 \rangle = \frac{1}{w_0^2}(2p + |\ell| +1)\;.
\end{equation}

\section{Appendix: Perfect Vortex Beams}
\label{app:PVB} For a realistic perfect vortex beam at $z = 0$, the radial profile is
given by Eq. \eqref{eq:PVB}.

For a sufficiently large argument $x \in \mathds{R}^+$,   the asymptotic expansion of the modified Bessel function is approximated by (see Eq. 8.451.5 of Ref. \cite{GR:2007}): 
\begin{equation}\label{eq:ModBessapprox}
    I_m(x) \approx \frac{e^x}{\sqrt{2\pi x}}\left[ 1 - \frac{4m^2-1}{8x} + ...\right]\;,
\end{equation}
Retaining only the leading term and setting $x = \frac{2Rr}{w^2}$, we obtain
\begin{equation}
    I_{\ell}\left(\frac{2Rr}{w^2}\right) \approx \frac{w}{2\sqrt{\pi R r}}e^{\frac{2Rr}{w^2}} \;.
\end{equation}
Substituting this approximation into  Eq. \eqref{eq:PVB} yields
\begin{equation}
     \mathcal{U}_\ell(r) \approx A\, \frac{w}{2\sqrt{\pi R r}} e^{-\frac{r^2 + R^2}{w^2}}e^{\frac{2Rr}{w^2}} = A \frac{w}{2\sqrt{\pi R r}}e^{-\frac{(r - R)^2}{w^2}}\;.
\end{equation}
Thus, the intensity profile forms a ring of radius $R$ and width $2w$. Let us define $\delta = \frac{R-r}{R}$, so that 
\begin{equation}
    \frac{1}{\sqrt{r}} = \frac{1}{\sqrt{R}}\frac{1}{(1-\delta)^{\frac{1}{2}}}\;.
\end{equation}
Since $R \gg w$, the field is only appreciable only in a narrow region around $r = R$. In this region, $\delta \ll 1$, and therefore, $\frac{1}{\sqrt{r}} \approx \frac{1}{\sqrt{R}}$. Under this approximation Eq. \eqref{eq:PVB} further reduces to 
\begin{equation}\label{eq:POV2}
    \mathcal{U}_{\ell,R}(r) \approx A\,  e^{-\frac{(r - R)^2}{w^2}}\;,
\end{equation}
It is important to note that the approximations are not valid for all values of $\ell$. Neglecting the second term in the asymptotic expansion given in Eq. \eqref{eq:ModBessapprox} requires $|\ell| \ll \frac{R}{w}$.

The following calculations rely on standard Gaussian integrals. Defining the error function:
\begin{equation}\label{eq:erf}
    \text{erf}(x) = \frac{2}{\sqrt{\pi}}\int_0^x e^{-t^2}\,dt\;, 
\end{equation}
the relevant integrals, valid for $q > 0$, are
\begin{align}
    &\int_0^\infty e^{-q^2x^2}\,dx = \frac{\sqrt{\pi}}{2q} \;, \label{eq:G5} \\
    &\int_0^u e^{-q^2x^2}\,dx = \frac{\sqrt{\pi}}{2q}\,\text{erf}(qu)\;, \label{eq:G1} 
\end{align}
\begin{align}
    &\int_0^\infty x\,e^{-q^2x^2}\,dx = \frac{1}{2q^2} \,, \label{eq:G6} \\
    &\int_0^u x\,e^{-q^2x^2}\,dx = \frac{1}{2q^2}\left[1 - e^{-q^2u^2} \right]\;, \label{eq:G2} 
\end{align}
\begin{align}
     &\int_0^\infty x^2\,e^{-q^2x^2}\,dx = \frac{\sqrt{\pi}}{4q^3}, \label{eq:G7} \\
    &\int_0^u x^2\,e^{-q^2x^2}\,dx = \frac{1}{2q^3}\left[\frac{\sqrt{\pi}}{2}\text{erf}(qu) - qu\,e^{-q^2u^2}  \right] \;, \label{eq:G3}
\end{align}
\begin{align}
    &\int_0^\infty x^3\,e^{-q^2x^2}\,dx = \frac{1}{2q^4} \label{eq:G8} \;,\\
    &\int_0^u x^3\,e^{-q^2x^2}\,dx = \frac{1}{2q^4} \left[1 - (1+ q^2u^2)\,e^{-q^2u^2} \right] \label{eq:G4} \;.
\end{align}

\subsection{Normalization Constant}

From the normalization condition of $\psi(r,\phi)$ [Eq. \eqref{eq:Norm}], and using the approximate form of the PVB given in \eqref{eq:POV2}, we obtain:
\begin{equation}
    \frac{1}{A^2} = 2\pi \int_0^\infty r e^{-2\frac{\left(r - R \right)^2}{w^2}} dr\;.
\end{equation}
Introducing the change of variable $v = r - R$, the resulting two integrals can be evaluated using expressions \eqref{eq:G5} - \eqref{eq:G2}, giving:
\begin{equation}\label{eq:A2}
    \frac{1}{A^2} = 2\pi \Biggl[ \frac{\sqrt{\pi}}{2\sqrt{2}}Rw \left(1 + \text{erf}\left(\frac{\sqrt{2}R}{w}\right)\right)+\frac{w^2}{4}e^{-\frac{2R^2}{w^2}}\Biggr]\;.
\end{equation}

Since $\frac{R}{w} \gg 1$, we may approximate $\text{erf}\left(\frac{\sqrt{2}R}{w}\right) \approx 1$ and $e^{-\frac{2R^2}{w^2}} \approx 0$, which results in
\begin{equation}\label{eq:A3}
    \frac{1}{A^2} \approx 2\pi \frac{\sqrt{\pi}}{\sqrt{2}}Rw\;.
\end{equation}

\subsection{Expectation Value of $x^2$}

Substituting the approximate PVB expression from Eq.\eqref{eq:POV2} into Eq.\eqref{x2}, we find
\begin{equation}\label{eq:x2IntPOV}
    \langle x^2 \rangle = \pi A^2\int_0^\infty r^3 e^{-\frac{2(r-R)^2}{w^2}}\,dr\;.
\end{equation}
Using the change of variable $v = r - R$:
\begin{equation}\label{eq:x^2POV1}
    \langle x^2 \rangle = \pi A^2\int_{-R}^\infty \left(v^3 +3Rv^2 + 3R^2v+R^3\right)e^{-\frac{2v^2}{w^2}}dv\;.
\end{equation}
Using results \eqref{eq:G5}-\eqref{eq:G4}, we obtain:
\begin{equation}
\begin{split}
    \langle x^2 \rangle =  \,\pi A^2 \Biggl[&\frac{\sqrt{\pi}Rw}{8\sqrt{2}}\left(1 + \text{erf}\left( \frac{\sqrt{2}R}{w}\right) \right)\left(3w^2 + 4R^2 \right)  \\&+ \frac{w^2}{8}e^{-\frac{2R^2}{w^2}}\left( 2R^2 +w^2  \right) \Biggr]\;.  
\end{split}
\end{equation}
In the regime $\frac{R}{w} \gg 1$, we may approximate this result as
\begin{equation}
    \langle x^2 \rangle \approx \pi A^2 \frac{\sqrt{\pi}\,Rw}{4\sqrt{2}}(4R^2 + 3w^2)\;.
\end{equation}
Finally, substituting $A^2 \approx \frac{1}{\sqrt{2\pi}\,\pi\,Rw}$ [Eq. \eqref{eq:A3}], we arrive at 
\begin{equation}\label{eq:x2PVB}
    \langle x^2 \rangle \approx \frac{1}{2} \left(R^2+\frac{3w^2}{4}\right)\;.
\end{equation}

\subsection{Expectation Value of $r$}

Using the approximate PVB form given in Eq. \eqref{eq:POV2}, the expectation value of $r$ can be written as
\begin{equation}\label{eq:rIntPOV}
    \langle r\rangle = 2\pi A^2 \int_0^\infty r^2 e^{-\frac{2(r-R)^2}{w^2}}dr\;.
\end{equation}
Making the substitution $v = r - R$, we obtain
\begin{equation}
    \langle r\rangle = 2\pi A^2 \int_{-R}^\infty \left(v^2 +2Rv + R^2 \right)e^{-\frac{2v^2}{w^2}}dv\;.
\end{equation}
Using Eqs. \eqref{eq:G5}-\eqref{eq:G3}, we find
\begin{equation}
\begin{split}
    \langle r\rangle = 2\pi A^2 \Biggl[&\frac{\sqrt{\pi}Rw}{2\sqrt{2}}\left(1 + \text{erf}\left(\frac{\sqrt{2}R}{w}\right)\right)\left(R+\frac{w^2}{4R}\right) \\ &+ \frac{w^2}{4}e^{-\frac{2R^2}{w^2}}R \Biggr]\;.
\end{split}
\end{equation}
Since $\frac{R}{w} \gg 1$, this expression can be approximated as
\begin{equation}
    \langle r\rangle \approx 2\pi A^2\frac{\sqrt{\pi}\,Rw}{\sqrt{2}}\left(R + \frac{w^2}{4R} \right)\;.
\end{equation}
Finally, inserting the previously determined normalization constant from Eq. \eqref{eq:A3}, we obtain:
\begin{equation}\label{eq:rPVB}
    \langle r\rangle \approx R + \frac{w^2}{4R}\;.
\end{equation}

\subsection{Expectation value of $k_x^2$}

Inserting the approximate PVB expression from Eq. \eqref{eq:POV2} into Eq. \eqref{eq:k_xk_xOAM} and using
\begin{equation}
    \frac{d\,\mathcal{U}(r)}{dr} = -\frac{2A}{w^2}(r-R)e^{-\frac{(r-R)^2}{w^2}}\;,
\end{equation}
we obtain
\begin{equation}
    \langle k_x^2 \rangle =  \pi A^2 \int_0^\infty \left(\frac{4r}{w^4} (r-R)^2 e^{-\frac{2(r-R)^2}{w^2}} + \frac{\ell^2}{r}e^{-\frac{2(r-R)^2}{w^2}}\right)dr\;.
\end{equation}
A comparison of this integral with those given in  Eqs. \eqref{eq:x2IntPOV} and \eqref{eq:rIntPOV}, together with
\begin{equation}
    \left\langle \frac{1}{r^2} \right\rangle = 2\pi \int_0^\infty \frac{1}{r}|\mathcal{U}(r)|^2\,dr\;, 
\end{equation}
allows us to express $\langle k_x^2 \rangle$ in terms of other expectation values:
\begin{equation}
    \langle k_x^2 \rangle = \frac{4}{w^4}\left(\langle x^2 \rangle -R\langle r\rangle + \frac{R^2}{2}  \right) +\frac{ \ell^2}{2} \left\langle \frac{1}{r^2} \right\rangle  \;.
\end{equation}
Using Eqs. \eqref{eq:x2PVB} and \eqref{eq:rPVB},this becomes: 
\begin{equation}
 \langle k_x^2 \rangle =    \frac{1}{2}\left( \frac{1}{w^2} + \ell^2 \left\langle \frac{1}{r^2} \right\rangle \right)\;.
\end{equation}

Introducing the parameter $\delta = \frac{R-r}{R}$,we write
\begin{equation}
    \frac{1}{r^2} = \frac{1}{R^2}\frac{1}{(1-\delta)^2}\;.
\end{equation}
The Maclaurin expansion of $\frac{1}{(1-\delta)^2}$ reads
\begin{equation}
    \frac{1}{(1-\delta)^2} = \sum_{n= 0}^\infty (n+1)\delta^{n} = 1 + 2\delta + 3\delta^2 +...\;.
\end{equation}
Because $R \gg w$, the dominant contributions to the integral arise from values of $r$ close to $R$, implying $\delta \ll 1$. For this reason, we can approximate $\frac{1}{r^2} \approx \frac{1}{R^2}$, and the expectation value becomes
\begin{equation}\label{eq:kx2PVB}
    \langle k_x^2 \rangle = \frac{1}{2}\left(\frac{1}{w^2}+\frac{\ell^2}{R^2}\right)\;.
\end{equation}

%%%%%%%%%%%%%
%%%%%%%%%%%%%%%%%%%
\section{{Appendix: On Perfect Vortex Beams as a mode basis}}
\label{sec:PVBbasis}

{The PVBs are distinct from other mode families (such as the $LGs$) in that they are described by the discrete OAM winding number $\ell$, and the continuous parameter $R$ relating to the radial ring radius.  Here we derive the conditions in which the PVBs can be considered as a mode basis.}

{For convenience, we remember from \eqref{eq:PVB} that the PVB modes are defined as}
\begin{equation}
\Phi_{\ell, R}(r, \phi) = C_0\, e^{i \ell \phi}
  \,e^{-\frac{r^{2}+R^{2}}{w^{2}}}\, I_{\ell}\!\left(\frac{2 R r}{w^{2}}\right),
\label{eq:PV_def}
\end{equation}
{The modes $\Phi_{\ell,R}$ defined in Eq.~\eqref{eq:PV_def} with prefactor $C_0$ are \emph{not} individually normalized. Their norm, obtained by applying Eq.~\eqref{eq:GR_radial} with $\beta=\gamma=2R/w^2$ and $\alpha=2/w^2$, is
\begin{equation}
\|\Phi_{\ell,R}\|^2 = \frac{1}{w^2}\,e^{-R^2/w^2}\,I_\ell\!\left(\frac{R^2}{w^2}\right),
\label{eq:PVB_norm}
\end{equation}
which depends on both $R$ and $\ell$. The value of $C_0$ is therefore \emph{not} the per-mode normalization constant~$A$ of Appendix~\ref{app:PVB}; instead it is fixed entirely by the resolution-of-identity condition (see below), yielding $C_0 = \sqrt{2/\pi}/w^2$.

The use of a uniform, $\ell$-independent prefactor $C_0$ is a structural necessity of the completeness calculation, not merely a choice of convention. To establish completeness one must evaluate the sum $\sum_\ell e^{i\ell(\phi-\phi')}I_\ell(rr'/w^2)$ in closed form using the Bessel generating function $\sum_\ell e^{i\ell\varphi}I_\ell(z)=e^{z\cos\varphi}$. This identity requires the coefficient multiplying each $I_\ell$ to be the same for all $\ell$. The exact per-mode normalization constant
\begin{equation}
N_\ell = \sqrt{\frac{2}{\pi w^2 e^{-R^2/w^2} I_\ell(R^2/w^2)}}
\label{eq:Nell}
\end{equation}
carries $\ell$-dependence through the factor $I_\ell(R^2/w^2)$, and its use in place of $C_0$ would render the sum over $\ell$ intractable. The constant $A$ of Appendix~\ref{app:PVB} is the large-$R/w$ approximation to $N_\ell$; in that regime $I_\ell(R^2/w^2)\approx e^{R^2/w^2}/\sqrt{2\pi R^2/w^2}$ is $\ell$-independent at leading order, and one recovers $A^2\approx 1/(\sqrt{2\pi}\,\pi\,Rw)$.
}

\subsection*{{Orthogonality and overlaps}}

{The inner product of two PVB modes is}
\begin{equation}
\langle \Phi_{\ell,R} \,|\, \Phi_{\ell',R'} \rangle
= \int_{0}^{2\pi}\!\!d\phi \int_{0}^{\infty}\!\! r\,dr\;
  \Phi_{\ell,R}^{*}(r,\phi)\,
  \Phi_{\ell',R'}(r,\phi).
\end{equation}
Substituting Eq.~\eqref{eq:PV_def} and separating angular and radial parts, the angular integral gives $2\pi\delta_{\ell\ell'}$, so modes with different
azimuthal index are exactly orthogonal. The radial integral is then
\begin{equation}
|C_0|^2e^{-\frac{R^{2}+R^{\prime 2}}{w^{2}}}
\int_{0}^{\infty} r\,e^{-\frac{2r^{2}}{w^{2}}}
  I_{\ell}\!\left(\frac{2Rr}{w^{2}}\right)
  I_{\ell}\!\left(\frac{2R'r}{w^{2}}\right)dr,
\label{eq:overlap_factored}
\end{equation}
{
which can be evaluated using the formula
\cite{GR:2007}}
\begin{equation}
\int_{0}^{\infty} r\,e^{-\alpha r^{2}}\,
  I_{\ell}(\beta r)\,I_{\ell}(\gamma r)\,dr
= \frac{1}{2\alpha}\,
  e^{\frac{\beta^{2}+\gamma^{2}}{4\alpha}}\,
  I_{\ell}\!\left(\frac{\beta\gamma}{2\alpha}\right),
\qquad 
\label{eq:GR_radial}
\end{equation}
with $\mathrm{Re}(\alpha)>0,\;\ell>-1$. 
{The constraint $\ell>-1$ can here be resolved using the reflection identity $I_{-n}(x)=I_n(x)$, which holds for all integer
$n$, such that the integrand for order $-|\ell|$ is identical to that for order
$+|\ell|$. We can define $\alpha = 2/w^2$, $\beta = 2R/w^2$, $\gamma = 2R'/w^2$, giving
$(\beta^2+\gamma^2)/4\alpha = (R^2+R'^2)/2w^2$ and
$\beta\gamma/2\alpha = RR'/w^2$.
The radial integral therefore equals
$(w^2/4)\,\exp[(R^2+R'^2)/2w^2]\,I_\ell(RR'/w^2)$.
Inserting this result and fixing $C_0 = \sqrt{2/\pi}/w^2$ (see below), so that
$\pi |C_0|^2 w^2/2 = 1/w^2$, we obtain the overlap}
\begin{equation}
\langle \Phi_{\ell,R} \,|\, \Phi_{\ell,R'} \rangle
= \frac{1}{w^{2}}\,
  e^{-\frac{R^{2}+R^{\prime 2}}{2w^{2}}}\,
  I_{\ell}\!\left(\frac{R R^{\prime}}{w^{2}}\right).
\label{eq:PVB_overlap}
\end{equation}
{This is a smooth, positive function of $R$ and $R'$ for all $R,R'\geq 0$:
modes of the same $\ell$ but different $R$ are \emph{not} orthogonal in general. However, it does not necessarily imply that the PVB modes do not form a basis, as is the case for the coherent states in quantum optics \cite{perelomov1986}.
It does mean, however, that if the PVB family turns out to span the  space, it does so
with redundancy rather than as a minimal basis.  We investigate this question
in the next subsection.}
\par

{Two further remarks on Eq.~\eqref{eq:PVB_overlap} are worth noting.
First, it has the same functional form as the PVB radial
profile~\eqref{eq:PV_def} itself. That is, with $r\to R$, $R\to R'$, and
$w^2\to w^2/2$, the overlap function is itself a PVB amplitude evaluated at
$r=R$.
Second, in the limit $w\to 0$, we have
$I_\ell(x)\sim e^x/\sqrt{2\pi x}$ for $x=RR'/w^2\to\infty$, which gives}
\begin{equation}
\langle \Phi_{\ell,R} \,|\, \Phi_{\ell,R'} \rangle
\xrightarrow{\;w\,\to \,0\;}
\frac{\delta(R-R')}{R},
\label{eq:overlap_limit}
\end{equation}
{where we used
$(1/w\sqrt{2\pi})\exp[-(R-R')^2/2w^2]\to\delta(R-R')$
and $\sqrt{RR'}\to R$ at $R'=R$.
With the natural measure $R\,dR$ on the radial space, this gives
$\int_0^\infty \langle \mathcal{U}_{\ell,R}|\mathcal{U}_{\ell,R'}\rangle
R'\,dR' \to 1$ as $w\to 0$: the modes become orthonormal with respect to the
measure $R\,dR$ only in this limit.}
\par
{Finally, we note that the overlap~\eqref{eq:PVB_overlap} depends on the envelope parameter $w$
in a physically transparent way.  In the limit $w\to 0$ the family approaches an
orthonormal set, while for large $w$ the PVB radial ring
is broad, the overlaps are strong, and the family is far from orthogonal.
This $w$-dependence is qualitatively different from the quantum-optical coherent states \cite{perelomov1986}, for which the
overlap $|\langle\alpha|\beta\rangle|^2 = e^{-|\alpha-\beta|^2}$ is independent
of any tunable parameter and never vanishes: the non-orthogonality of coherent
states is irreducible, whereas for PVBs it is controlled by a physical parameter
--- the ring width --- and can be made arbitrarily small.}
\subsection*{Completeness}

{We now evaluate}
\begin{equation}
K=\sum_{\ell=-\infty}^{\infty} \int_{0}^{\infty} R\, dR\;
\Phi_{\ell, R}(r, \phi)\,
\Phi_{\ell, R}^{*}(r', \phi'),
\label{eq:completeness}
\end{equation}
which should give 
\begin{equation}
K=\delta^{(2)}(\mathbf{r}-\mathbf{r}'),
\label{eq:completenessK}
\end{equation}
 if the PVBs form a complete basis.
Writing out the left-hand side and separating the $R$-dependence:
\begin{align} \label{eq:completeness_lhs}
K
= & |C_0|^{2}\,e^{-\frac{r^{2}+r^{\prime 2}}{w^{2}}}
  \sum_{\ell} e^{i\ell(\phi-\phi')} \times \\ \nonumber
 &  \int_{0}^{\infty} R\,e^{-\frac{2R^{2}}{w^{2}}}
I_{\ell}\!\left(\frac{2Rr}{w^{2}}\right) I_{\ell}\!\left(\frac{2Rr'}{w^{2}}\right)dR.
\end{align}
{The $R$-integral has {exactly} the same form as the $r$-integral in the
overlap calculation, with $r\leftrightarrow R$, $R\leftrightarrow r$, and
$R'\leftrightarrow r'$.  Applying Eq.~\eqref{eq:GR_radial} with
$\alpha=2/w^2$, $\beta=2r/w^2$, $\gamma=2r'/w^2$ gives}
\begin{equation}
 \frac{w^2}{4}\,e^{\frac{r^2+r'^2}{2w^2}}\,
  I_{\ell}\!\left(\frac{rr'}{w^{2}}\right).
\label{eq:R_integral}
\end{equation}
{Substituting this into Eq.~\eqref{eq:completeness_lhs}, we have}
\begin{equation}
K
= \frac{|C_0|^{2} w^2}{4}\,
  e^{-\frac{r^{2}+r^{\prime 2}}{2w^{2}}}
  \sum_{\ell} e^{i\ell(\phi-\phi')}\,
  I_{\ell}\!\left(\frac{rr'}{w^{2}}\right).
\label{eq:completeness_mid}
\end{equation}
{The remaining sum over $\ell$ can be evaluated using the modified Bessel generating
function $\sum_{\ell=-\infty}^{\infty} e^{i\ell\varphi}\,I_\ell(z) = e^{z\cos\varphi}$ \cite{GR:2007}.
With $z=rr'/w^2$ and $\varphi=\phi-\phi'$, we have}
\begin{equation}
\sum_{\ell} e^{i\ell(\phi-\phi')}\,I_{\ell}\!\left(\frac{rr'}{w^2}\right)
= \exp\!\left(\frac{rr'\cos(\phi-\phi')}{w^2}\right).
\label{eq:bessel_gen}
\end{equation}
{Inserting Eq.~\eqref{eq:bessel_gen} into Eq.~\eqref{eq:completeness_mid} and
using $r^2+r'^2-2rr'\cos(\phi-\phi')=|\mathbf{r}-\mathbf{r}'|^2$, we arrive at}
\begin{equation}
K
= \frac{|C_0|^{2} w^2}{4}\,
  e^{-\frac{|\mathbf{r}-\mathbf{r}'|^{2}}{2w^{2}}}
= G_{w}(\mathbf{r}-\mathbf{r}'),
\label{eq:completeness_gaussian}
\end{equation}
where the last equality fixes the normalisation constant via
$|C_0|^2 w^2/4 = 1/(2\pi w^2)$, i.e.\
$C_0 = \sqrt{2/\pi}/w^2$, and
\begin{equation}
G_{w}(\mathbf{r}-\mathbf{r}') \equiv
\frac{1}{2\pi w^2}\,
e^{-\frac{|\mathbf{r}-\mathbf{r}'|^{2}}{2w^{2}}}
\label{eq:heat_kernel}
\end{equation}
is a normalised 2D Gaussian kernel.
{The sequence $\{G_w\}_{w>0}$ satisfies the standard criteria for a Dirac delta sequence \cite{arfken2013}.
}
{Therefore $G_w\to\delta^{(2)}(\mathbf{r}-\mathbf{r}')$ distributionally as $w\to 0$, which implies that we can establish Eq.~\eqref{eq:completenessK} only in that limit.} This is physically natural: sharper ring profiles ($w\to 0$)
resolve finer radial structure, and in this limit the PVB family saturates the full
$L^2(\mathbb{R}^2)$ space.

{\subsection*{Practical validity of the PVB expansion}}
{In practice, when $w$ is finite, the completeness of the PVB ``basis"
is not exact but approximate.  Let us briefly discuss this fact and determine in what conditions a coherent field can be expanded in terms of PVBs. The PVB
expansion of any field $\psi(\mathbf{r})$ is:}
\begin{equation}
\sum_{\ell=-\infty}^{\infty}\int_{0}^{\infty} R\,dR\;
  c_{\ell,R}\;\Phi_{\ell,R}(\mathbf{r})
= \int G_{w}(\mathbf{r}-\mathbf{r}')\,\psi(\mathbf{r}')\,d^{2}r',
\label{eq:PVB_reconstruction}
\end{equation}
{where $c_{\ell,R}=\langle\Phi_{\ell,R}|\psi\rangle$ and $G_{w}$ is defined in Eq.~\eqref{eq:heat_kernel}. This is not $\psi$ itself but its
convolution with the Gaussian kernel $G_{w}$, which acts as a spatial filter.
In Fourier space, convolution with $G_{w}$ multiplies each spatial-frequency
component by the transfer function $\widetilde{G}_{w}(\mathbf{k})=e^{-w^2 k^2/2}$,
which suppresses transverse wavenumbers $k\gtrsim 1/w$.  The PVB expansion
is therefore a faithful representation of $\psi$ provided $\psi$ contains no
significant structure on transverse scales smaller than $w$, i.e.\ provided}
\begin{equation}
w \ll \lambda_{\min}(\psi),
\label{eq:PVB_condition}
\end{equation}
{where $\lambda_{\min}(\psi)$ denotes the shortest transverse length scale present
in $\psi$.}
\par
{{We note from \eqref{eq:PVB_reconstruction} that the OAM content of $\psi$ is always faithfully represented.}
This follows from the fact that the kernel $G_{w}(\mathbf{r}-\mathbf{r}')$ is isotropic.  In Fourier space, the isotropic filter $e^{-w^2k^2/2}$ does
not mix azimuthal harmonics.  Therefore the OAM spectrum of $\psi$ is reproduced exactly
by the PVB expansion for any $w>0$: the angular part of the resolution of identity is
exact (it follows from Fourier completeness on the circle, not from the $w\to 0$
limit), and only the radial profile is affected by the finite-$w$ smoothing.}

{Let us illustrate the discussion above by considering the PVB expansion of an LG mode.
Consider $\psi=\operatorname{LG}_{p,\ell_0}$ with beam waist $w_\psi$.  Its OAM index
$\ell_0$ is preserved exactly as noted above.  Its radial profile is governed by the
associated Laguerre polynomial $L_p^{|\ell_0|}(2r^2/w_\psi^2)$, which has $p$ radial
nodes. {For large $p$, the nodes are approximately separated by $\Delta r \sim w_\psi/\sqrt{2p}$, following the asymptotic spacing of Laguerre polynomial zeros,} so the finest radial scale is $\lambda_{\min}=w_\psi/\sqrt{2p}$.
%For $p>0$ the nodes are spaced approximately $\Delta r\sim w_\psi/\sqrt{2p}$ apart (from the asymptotic spacing of Laguerre polynomial zeros for large $p$), so the finest radial scale is $\lambda_{\min}=w_\psi/\sqrt{2p}$.
The convolution
$(G_{w}*\operatorname{LG}_{p,\ell_0})$ smooths over these nodes and the
reconstruction degrades once $w\gtrsim w_\psi/\sqrt{2p}$.  The condition for a
faithful expansion is therefore}
\begin{equation}
w \;\ll\; \frac{w_\psi}{\sqrt{2p}}\,,
\label{eq:LG_condition}
\end{equation}
{which reduces to $w\ll w_\psi$ for the pure-vortex case $p=0$, in which no radial
nodes are present and only the overall beam scale must be resolved.
Equation~\eqref{eq:LG_condition} shows that higher radial order demands
proportionally sharper PVB rings. In the opposite limit $w\gg w_\psi/\sqrt{2p}$ the
reconstructed field retains the correct OAM but its radial structure is washed out,
leaving only the envelope $e^{-r^2/w_\psi^2}$ modulated by $e^{i\ell_0\theta}$.}

\subsection*{{Analytic expansion coefficients for $p=0$ LG modes}}

{For the pure-vortex case $p=0$, the expansion coefficients $c_{\ell,R}=\langle\Phi_{\ell,R}|\psi\rangle$ with $\psi=\operatorname{LG}_{0,\ell_0}$ can be evaluated in closed form. Restricting to $\ell=\ell_0$ (the only non-vanishing azimuthal sector, by the angular orthogonality established above), we insert the LG mode
\begin{equation}
\mathcal{U}_{0,\ell_0}(r) = \sqrt{\frac{2}{\pi\,|\ell_0|!}}\,\frac{1}{w_\psi}\left(\frac{\sqrt{2}\,r}{w_\psi}\right)^{|\ell_0|} e^{-r^2/w_\psi^2}
\end{equation}
into the overlap integral and perform the radial integration using  Eq.~6.631-1 of Ref. \cite{GR:2007},
\begin{equation}
\int_0^\infty x^{\nu+1}e^{-\alpha x^2}I_\nu(\beta x)\,dx = \frac{\beta^\nu}{(2\alpha)^{\nu+1}}\exp\!\left(\frac{\beta^2}{4\alpha}\right),
\label{eq:GR6631}
\end{equation}
valid for $\mathrm{Re}(\alpha)>0$, $\mathrm{Re}(\nu)>-1$. Setting $\nu=|\ell_0|$, $\alpha=(w_\psi^{-2}+w^{-2})$, and $\beta=2R/w^2$, and collecting all prefactors, one obtains
\begin{equation}
c_{\ell_0,R} = N_{\ell_0}\,\frac{2^{|\ell_0|/2}}{\sqrt{\pi\,|\ell_0|!}}\,\frac{R^{|\ell_0|}\,w_\psi^{|\ell_0|+1}\,w^2}{\left(w_\psi^2+w^2\right)^{|\ell_0|+1}}\exp\!\left(-\frac{R^2}{w_\psi^2+w^2}\right),
\label{eq:cLR}
\end{equation}
where the per-mode normalization constant $N_{\ell_0}$ is given by Eq.~\eqref{eq:Nell}. In the regime $R\gg w$ the factor $N_{\ell_0}\approx(\sqrt{2\pi}\,R\,w)^{1/2}/w$ (from the asymptotic form of $I_{\ell_0}$) becomes $\ell_0$-independent, and Eq.~\eqref{eq:cLR} simplifies accordingly.

A few features of Eq.~\eqref{eq:cLR} are worth noting. First, the coefficient is maximized at $R^\star=\sqrt{(|\ell_0|+1)(w_\psi^2+w^2)/2}$, showing that the PVB ring radius that captures most of the LG mode's radial weight grows with OAM order as $R^\star\propto\sqrt{|\ell_0|}$, consistent with the known scaling of the LG peak radius. Second, the Gaussian factor $\exp(-R^2/(w_\psi^2+w^2))$ provides exponential suppression for rings much larger than $\sqrt{w_\psi^2+w^2}$, confirming that only a finite band of ring radii contributes significantly to the expansion. Third, Eq.~\eqref{eq:cLR} applies only to the $p=0$ case; for $p\geq 1$, the associated Laguerre polynomial introduces additional radial structure and the integral does not reduce to a single closed-form expression of this type.}
%%%%%%%%%%%%%
\section{Appendix: Bessel-Gaussian Beams}
\label{app:BG}
The radial component of a BG beam is given by Eq. \eqref{BG}.

\subsection{Fourier-Transform of a BG Beam}

Substituting the expression for the BG mode given in \eqref{BG} into Eq. \eqref{eq:Hankel} gives 
\begin{equation}
    \mathcal{H}_\ell(k_r) = B \int_0^\infty J_\ell (k_br)\,J_\ell(k_rr)\,e^{-r^2 / w_b^2}\, r\,dr\;.
\end{equation}
For $\ell < 0$, we use the relation $J_{-m}(x) = (-1)^m J_m(x)$ to write the integral as 
\begin{equation}
    \mathcal{H}_\ell(k_r) = B \int_0^\infty J_{|\ell|} (rk_b)\,J_{|\ell|}(k_rr)\,e^{-r^2 / w_b^2} \,r\,dr\;.
\end{equation}
This integral can be evaluated using Eq. 6.633.2 of Ref. \cite{GR:2007}, yielding
\begin{equation}
    \mathcal{H}_\ell(k_r) = B \frac{w_b^2}{2}e^{-\frac{\left(k_r^2+k_b^2\right)w_b^2}{4}}I_{|\ell|}\left(\frac{k_rk_bw_b^2}{2}\right)\;.
\end{equation}
Finally, inserting this expression into Eq. \eqref{FT}, we find that the Fourier transform of a BG beam is
\begin{equation}
    F_\ell(k_r,\theta) = (-i)^\ell B \,e^{i\ell\theta}\frac{w_b^2}{2}e^{-\frac{\left(k_r^2+k_b^2\right)w_b^2}{4}}I_\ell\left(\frac{k_rk_bw_b^2}{2}\right)\;.
\end{equation}
Comparing this result with the PVB expression in Eq. \eqref{PVB}, we see that the Fourier Transform of a BG beam corresponds to a PVB with parameters $R = k_b$ and $w = \frac{2}{w_b}$. Therefore, there is a direct correspondence between their second moments. The results for BG beams can be obtained from those of the PVBs by applying the substitution $x \leftrightarrow k_x$ with the appropriate change of variables. The regime $R\gg w$ and $|\ell|\ll \frac{R}{w}$, used in the PVB calculations, corresponds here to $w_b\gg \frac{1}{k_b}$ and $|\ell| \ll \frac{k_bw_b}{2}$.

\subsection{Expectation value of $x^2$}

The expectation value of $x^2$ can be obtained from the correspondence
 $\langle k_x^2 \rangle_{\text{PVB}}\leftrightarrow \langle x^2 \rangle_{\text{BG}}$. Using Eq. \eqref{eq:kx2PVB} and making the substitutions $R \rightarrow k_b$ and $w \rightarrow \frac{2}{w_b}$, we obtain
\begin{equation}\label{eq:x2BG}
    \langle x^2\rangle = \frac{1}{2}\left(\frac{w_b^2}{4}+\frac{\ell^2}{k_b^2} \right)\;.
\end{equation}

\subsection{Expectation value of $k_x^2$}

Similarly to $\langle x^2\rangle$, the expectation value of $k_x^2$ follows from the correspondence $\langle x^2 \rangle_{\text{PVB}}\leftrightarrow \langle k_x^2 \rangle_{\text{BG}} $. Applying the substitutions $R \rightarrow k_b$ and $w \rightarrow \frac{2}{w_b}$ to Eq.\eqref{eq:x2PVB} leads to
\begin{equation}\label{eq:kx2BG}
    \langle k_x^2\rangle = \frac{1}{2}\left(k_b^2+\frac{3}{w_b^2} \right)\;.
\end{equation}
%\spw{This result follows correctly from the substitution $w\to 2/w_b$ applied to $\langle x^2\rangle_\text{PVB} = \frac{1}{2}(R^2+\frac{3w^2}{4})$, giving $\frac{3w^2}{4}\to\frac{3}{w_b^2}$. However, the main-text Eq.~\eqref{eq:kr2BG} states $\langle k_r^2\rangle_\text{BG}=k_b^2+\frac{3}{2w_b^2}$, which would imply $\langle k_x^2\rangle=\frac{1}{2}(k_b^2+\frac{3}{2w_b^2})$. These two results are inconsistent. Please verify by direct calculation which is correct, and correct the other.}

%%%%%%%%%%%%%%%%%%%

%%%%%%%%%%%%%%%%%%%%%%%%%%%%%%%%%
\section*{Ethics Statement}
This article does not contain any studies involving human participants or animals performed by any of the authors.

\begin{acknowledgments}
    This research was funded by Fondo Nacional de Desarrollo Científico y Tecnológico (FONDECYT) Regular Grants No. (1240746, 1230796, 1231940, 1260111), ANID -- Millennium Science Initiative Program -- ICN17$_-$012,  ANID Anillo Project ATE250003, and ANID Doctoral Fellowship Grant No. 21241835.
\end{acknowledgments}

%\bibliographystyle{apsrev}
%\bibliography{CM_cylindrical}

  %%%%%%%%%%%%%%%%%%%%%%%%%%%%%%%%%%%%%%%%%%%%%%%%%%%%%%%%%%

\end{document}